\documentclass{aa}  

\usepackage{graphicx}
\usepackage{txfonts}
\begin{document} 

   \title{A light strange star in the remnant HESS J1731-347: minimal consistency checks }

   \author{J.E. Horvath
          \inst{1},
          L.S. Rocha
          \inst{1}.
          L.M. de S\'a
          \inst{1},
          P.H.R.S. Moraes
         \inst{2},
         L. G. Bar\~ao
          \inst{1},
         M.G.B. de Avellar
         \inst{1},
          A. Bernardo
          \inst{1},
          \and
           R.R.A. Bachega
         \inst{1},
         \fnmsep
     }

\institute{Universidade de S\~ao Paulo, Instituto de Astronomia, Geof\'i sica e Ci\^encias Atmosf\'ericas\\
              R. do Mat\~ao, 1226, Cidade Universit\'aria, 05508-090 S\~ao Paulo SP, Brazil\\    \email{foton@iag.usp.br}     
         \and
             Universidade Federal do ABC (UFABC), Centro de Ci\^encias Naturais e Humanas (CCNH)\\
Avenida dos Estados, 5001, 09210-580, Santo Andr\'e - SP, Brazil\\
             }

   \date{}

% \abstract{}{}{}{}{} 
% 5 {} token are mandatory
 
  \abstract
  % context heading (optional)
  % {} leave it empty if necessary  
   {Recently, Doroshenko and collaborators reported a very low-mass compact star, a Central Compact Object named XMMU J173203.3-344518 inside the supernova remnant HESS J1731-347. Its tiny mass is at odds with all calculations of minimum masses of neutron stars generated by iron cores, therefore (and even if not compellingly) it has been suggested to be a {\it strange star}.
   Besides the mass, a radius and surface temperature have been extracted from data, and the whole body of information should ultimately reveal if this object is truly consistent with an exotic composition.}
  % aims heading (mandatory)
   {To understand the status of the compact object XMMU J173203.3-344518 in HESS J1731-347 within the existing models of strange stars, including its prompt formation.}
  % methods heading (mandatory)
   {The information obtained on the mass, radius and surface temperature are compared to theoretical calculations performed within usual theoretical models using General Relativity as the assumed theory of gravitation and a handful of cooling scenarios. A qualitative discussion showing the consistency of the strange-matter driven supernova scenario with a low-mass compact star is provided.}
  % results heading (mandatory)
   {We found that the object HESS J1731-347 fits within the same quark star models recently employed to explain the masses and radii of the NICER objects PSR J040+6620 and PSR J0030+0451, in which both quantities were simultaneously determined. It is also remarkable to find that a simple cooling scenario devised $30\text{ yr}$ ago with superconducting quarks provides an overall good explanation of the surface temperature.}
  % conclusions heading (optional), leave it empty if necessary 
   {We conclude that XMMU J173203.3-344518 in the remnant HESS J1731-347 fits into a ``strange star'' scenario that is also consistent with heavier compact stars, which can also belong to the same class and constitute an homogeneous type of self-bound objects produced in Nature.}

\titlerunning{A light strange star in HESS J1731-347}
\authorrunning{J.E. Horvath et al.}
\maketitle
\section{Introduction}

The quest for the state of matter in its ultimate dense version (inside compact stars) has been going on for decades. Since it was clear from the early days that densities above the saturation value $\rho_{0} \sim 2.4 \times 10^{14}\text{ g cm}^{-3}$ were involved, all sorts of approaches have been devised to reliably describe the behavior of degrees of freedom in that regime. However, and in spite of abundant data from heavy-ion colliders \citep{Elena}, and extensive work on compact star physics \citep{Frido}, we still have only a basic knowledge about matter under these extreme conditions. 

One of the most fundamental questions is the very type of degrees of freedom present at the highest densities. Work on known nuclear species (hyperons, resonances, etc.) occupies most of the literature, but also the {\it exotic} composition has been 
amply explored. In addition to the liberation of quarks and gluons degrees of freedom in heavy ion collisions \citep{Larry}, 
quark matter has been explored indirectly as an important potential component of compact star composition \citep{Ivanenko}. In the most extreme proposal, a cold variety of this matter was suggested to be absolutely stable, and therefore a candidate to compose all but the outermost crust of a compact star \citep{Bodmer, Terazawa, Itoh, Witten}, and because of the large abundance and fundamental role of the $s$ quark in the mixture, called {\it strange quark matter} (SQM).

The search for evidence of a SQM composition of compact stars has been extensive, with some reports and arguments being withdrawn, and others still being considered (for example, \citet{DiS}). 
The recent report by \citet{Doro} of a low mass for the Compact Central Object XMMU J173203.3-344518 in the remnant HESS J1731-347 adds interest to this quest. The observations 
revealed a mass below the $1 \, M_{\odot}$ scale, the formation of which is thought to be not viable in standard Stellar Evolution \citep{japas}. The measurement of the radius is also interesting, because it came out slightly smaller than the most reliable currently known ones. This was possible by means of a spectral modeling and the use of {\it Gaia} data to refine the distance estimate. Therefore, this compact object became a candidate to a strange star, as stated by the authors, although a ``normal'' composition is by no means excluded. We shall show below that the XMMU J173203.3-344518 candidate passes the minimal requirements to be considered a strange star, and discuss possible reasons for its very low mass. 
  
\section{Mass and radius of XMMU J173203.3-344518 in HESS J1731-347 within existing models}

A whole class of quark matter models were investigated in the last $20\text{ yr}$ or so, after the recognition of the importance of quark-quark paring in dense matter. The pairing was perturbatively discussed earlier \citep{BL}, but the new calculations were performed within a non-perturbative approach and 
suggested a much larger value of the gap $\Delta$ \citep{Alf}. For large densities the so-called Color-Flavor Locked phase (CFL) in which all quarks pair, discussed in \citet{Alf} is expected to be the ground state. Even though the densities inside compact stars should not be so extreme, the asymptotic CFL description was adopted as a plausible model. In fact, the pairing has been shown to considerably enlarge the ``stability window'' \citep{LH1} for CFL strange matter. Other pairing patterns, for example, the 2SC that allows only the lighter quarks to pair, have been explored \cite{Alf, Krishna} and may be relevant if the actual density inside compact objects is not too high. Ultimately, these theoretical possibilities must be related to actual observations.

The equation of state (EoS) including pairing effects has been discussed in the literature. In the case of the CFL state, the simplest version of it \citep{Alf, LH1} reads

\begin{equation}
\label{eq:pressure}
    P_r = \frac{1}{3}\rho + \frac{2\Phi}{\pi}\rho^{1/2} - \left( \frac{3\Phi^2}{\pi^2} + \frac{4}{3}B \right),
\end{equation}

\noindent with $\Phi$ defined by

\begin{equation}
\label{eq:phi}
    \Phi = -\frac{m_s^2}{6}+\frac{2\Delta^2}{3},
\end{equation}

\noindent where $m_s$ is the mass of the strange quark, $\Delta$ the pairing gap (usually treated as a parameter) and $B$ is the bag constant. In the limit $m_s \rightarrow 0$ and non-interacting (unpaired) quarks, a \textit{MIT bag model}-like EoS is restored \citep{AFO}. 

Variants of this self-bound EoS have been considered over the years. In a recent work \citep{HM} we have collected a few examples that allow an explanation of the masses and radius of the most reliable simultaneous measurements available, those presented in \cite{NICER1} for PSR J0030+0451 and \cite{NICER2} for PSR J0740+6620. The compact object in the GW merger GW190814 \citep{GW} (although no estimate for the radius is available) is a candidate to the heaviest neutron star known \citep{nos}, and the stellar sequences may be arguably required to go up to this high value. It is important to remark that the analysis of the existing sample of galactic neutron stars \citep{nos} favors a large maximum mass, and also that a similar value for it was found in an analysis of the GW sample \citep{GWM}. The maximum mass should be above $2.2 \, M_{\odot}$ empirically \citep{Romani, GWM, nos}.

The four EoS employed in \citet{HM} correspond to self-bound versions of cold quark matter. The first is the NJL-CFL EoS with vector interactions \citep{Ferrer} in which the self-bound condition is achieved by introducing a vacuum term and repulsive vector interactions boost the maximum mass of the stellar sequence upwards. The Mean Field Theory of QCD (MFTQCD) \citep{Franzon} with the parameter $\xi$ in the range of self-boundness, and two versions of the CFL model with a self-bound EoS given in Eqs. \eqref{eq:pressure} and \eqref{eq:phi}, solved within an exact anisotropic metric {\it ansatz} due to Thirukkanesh-Ragel-Malaver \citep{Tikka, Mala}. This last model was analytically solved in \citet{Rocha}. 

Since we have derived in \citet{Rocha} an exact solution, some features of the 
models can be ``read off'' quite simply. For example, the two anisotropic models presented in \citet{HM} were calculated using a large value of the pairing gap 
$\Delta = 100 \, MeV$. Given that cooling suggests a very small gap (see below), this may be seen as misleading. However, from the anisotropic analytic solution we can 
readily check that even in the limit of ``zero'' pairing, the sequences display a lower maximum mass than the ones corresponding to large $\Delta$, but not dramatically so. The curves are not too different mainly because of the presence of the first term $m_{s}^{2}$ in the quantity $\Phi$ of eq.(2), which precludes a dramatic drop of $M_{max}$ in the $\Delta \rightarrow 0$ limit. As a consequence, low-gap anisotropic CFL would be viable models for all the objects addressed here. We have added one of such models (CFL with "zero gap" in the anisotropic \textit{ansatz}) in Fig. 1 for comparison.

All these self-bound EoS, with a zero-point pressure at finite density, produced the sequences of quark stars seen in Fig. 1. Without any change, we have plotted the new object HESS J1731-347 ($M = 0.77^{+0.20}_{-0.17} \, M_{\odot}$ and $R = 10.4^{+0.86}_{-0.78}\text{ km}$) considering a rectangle of $1 \sigma$ and the result shows that the same description attempted for the heavier objects is a very good model for the low-mass compact star reported by \citet{Doro}. Larger contours, beyond $1 \sigma$, would make the agreement even more plausible for the HESS object and the three other cases. Therefore, we consider this consistency check as successful. 

\begin{figure}
\resizebox{\hsize}{!}{\includegraphics{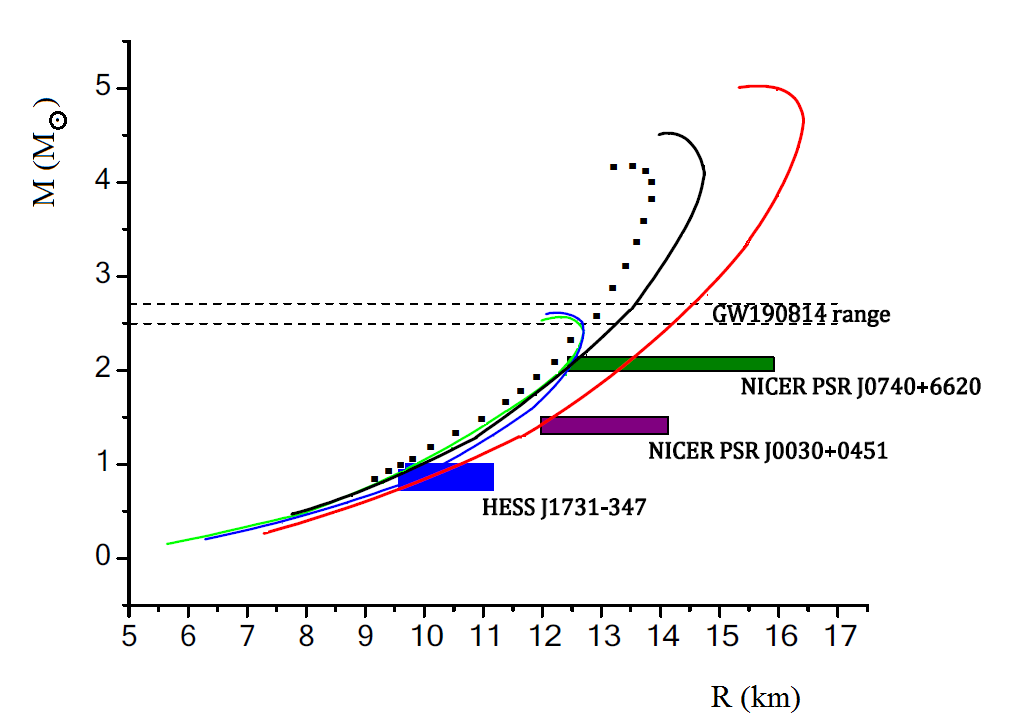}}
\caption{Mass-radius relation of the theoretical models vs. observed data. The green curve corresponds to the MFTQCD quark EoS, with $\xi = 0.003658\text{ MeV}^{-1}$ and $B_{QCD} = 62\text{ MeV/fm}^{3}$. The blue curve depicts the results for the NJL CFL+vector interactions EoS with $B_{0} = 57.3\text{ MeV/fm}^{3}$ and $G_{V}/G_{S} = 0.5$. Two anisotropic models corresponding to the Thirukkanesh-Ragel-Malaver {\it ansatz}, both with $\Delta = 100~\mathrm{MeV}$ and $m_s = 150~\mathrm{MeV}$ appear in red (corresponding to $B = 57.5~\mathrm{MeV/fm^3}$) and black ($B = 70~\mathrm{MeV/fm^3}$). The dotted sequence is the $\Delta \rightarrow 0$ for the same anisotropic models with $B = 70~\mathrm{MeV/fm^3}$, showing that the downward shift in the stellar sequence curves is not dramatic and may be viable as models of the new object. Data rectangles are $1 \sigma$ estimates for PSR J0030+0451 (purple), PSR 0740+6620 (green) and the new reported HESS J1731-347 (blue).}
\label{}
\end{figure}

\section{Surface temperature of XMMU J173203.3-344518 in HESS J1731-347}

Before the announcement of a low-mass value by \citet{Doro}, the central object was known with the name XMMU J173203.3-344518 \citep{Hess}, although there was considerable doubt concerning its age. The high temperature reported by \citet{Hess}, $T_{e}^{\infty} = 153^{+4}_{-2}\text{ keV}$, was difficult to reconcile with a $27\text{ kyr}$ age, although later a much younger age was found for the shorter distance of the admissible range \citep{Acero, Cui, Maxted}, which is now believed to lie in the range $2-6 \, kyr$.

The thermal history of compact stars has been modelled for many years (see, for example, \citet{Ken}). In its basic form (ignoring gravity for now), the cooling equation is just

\begin{equation}
\label{eq:cooling}
    C_{V} {\frac{dT}{dt}} = -\sum_i L_{i},
\end{equation}

\noindent where $C_{V}$ is the heat capacity, dominated by the quark component, and $L_{i}$ are the luminosities resulting from the processes that produce neutrinos and photons. 
The early calculations of quark matter recognized that a rapid cooling would ensue, when compared to hadronic compositions. However, the pairing of quarks due to attractive interactions, recognized in the perturbative calculations in the 1980's \citep{BL}, and later in a non-perturbative fashion \citep{Alf}, changed things substantially. As stated in \citet{Dany}, CFL pairing with large gaps would render paired quarks essentially passive, since both $C_{V}$ and $L_{\nu}$ would be strongly suppressed on both sides of Eq. \eqref{eq:cooling} for temperatures much lower than the gap scale $\sim 100\text{ MeV}$. However, if the pairing gap is much lower, of the order of the temperature 
$\Delta \leq 1\text{ MeV}$, the cooling can mimic the slow history of hadronic composition, and the large drop in $T_{e}^{\infty}$ does not occur promptly. 

\begin{figure}
\resizebox{\hsize}{!}{\includegraphics{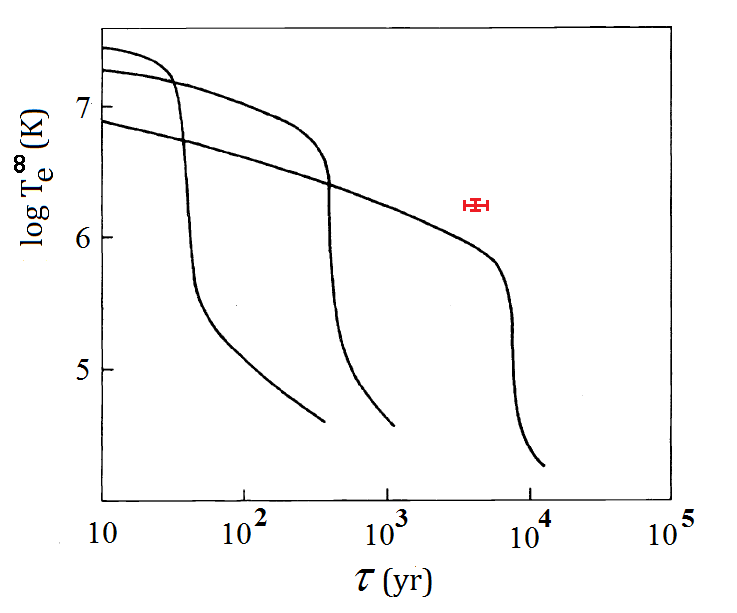}}
\caption{Surface temperature of HESS J1731-347 \citep{Hess}, marked with the red cross, contrasted with the early theoretical models cooling with paired quarks \citep{coolNOS} with small $\Delta$ gaps, for three different initial temperatures. Refined scenarios 
\citep{David} confirm the relatively slow cooling, instead of a large drop 
in $T_{e}^{\infty}$, for up to $\sim 10^{5}\text{ yr}$ provided the gap is small (see text)}
\label{CoolSS}
\end{figure}

The first attempt to model the latter behavior \citep{coolNOS} is useful to illustrate the situation, as shown in Fig. \ref{CoolSS}. The exact numerical values of the theoretical curves should not be taken too seriously, because many details were oversimplified or ignored in 1991. However, the comparison is useful to state the consistency of the surface temperature with the structure considerations derived from self-bound EoS, provided the pairing gap is {\it much smaller} than the $10-100\text{ MeV}$ ballpark considered in CFL models today. One of the suitable patterns of pairing discussed over the years is the 2SC \cite{David, David1, Armen, Bede, Noda}, expected to have small gaps and to be relevant at densities which are not extreme.The specific case of cooling, linked to a low mass which prompted a strange star suggestion, may be a good clue to go forward in these issues, stimulating further calculations. Let us remark that {\it rapid} cooling was sought as a signal of exotic composition over the years, whereas we now have a hint that the opposite may be true, as shown in the first papers addressing paired quark matter \citep{coolNOS}. 

\section{Formation scenario for HESS J1731-347: why is the mass so low?}

The conversion of neutron matter into strange quark matter (SQM) is now a venerable subject, started before 1990. While the first attempts to model the propagation of the conversion from neutrons to SQM were made within a laminar deflagration framework \citep{Angela, Jes}, it soon became clear that a much faster combustion mode should be considered \citep{HB}, prompted by the action of instabilities. Both Landau-Darreius and Rayleigh-Taylor instabilities quickly wrinkle the flame front within $\sim\text{ms}$ and set the stage for an acceleration. The burning enters the so-called {\it flamelet} regime, in 
which the flame propagation is still determined by diffusion, but the total burning rate is in turn controlled by turbulence, in a turbulent region called the {\it flame brush}. In the 
flamelet regime, for all scale $\gg l_{G}$ (the {\it Gibson scale}) \citep{turb}, the  turbulent velocity $u_{turb}$ and front width $l_{turb}$ are determined by the turbulence spectrum at the largest scales (very likely of a Kolmogorov type). From there, the possibility of the combustion entering the {\it distributed} regime, in which turbulent eddies disrupt the flame and dominate the burning on macroscopic and microscopic scales is substantial. In \citet{turb} the conditions for the combustion to ``jump'' to the detonation branch were discussed, although the very mechanism behind such transitions is poorly known and may be in fact multiple \citep{Shoichi}. Dedicated simulations have not produced a transition to a detonation stage \citep{Shoichi, Rachid}, although this does not seem definitive.

In the general case, the hypothesis that supernovae are aided by SQM appearance \citep{Macacos} helps to visualize that the supernova event provides a clue for its reported low mass. The suggestion is that the burning of neutrons $n \rightarrow \text{SQM}$ has enough energy to blow out a substantial mass from the outer core, provided the conversion happens in the detonation mode. This is true even for cores {\it more massive} than the minima of \citet{japas}, because their density profile does not need to be sharply decreasing \citep{Adam}. It is not known how the details would work, because a numerical simulation of the process is not available for these events, which would be the common case in Nature. The possibility that the progenitor was a low mass star in the Main Sequence, even one undergoing an electron-capture supernova first (and {\it not} an iron core collapse), would be also possible, since the $\text{O}-\text{Mg}-\text{Ne}$ white dwarf-like core collapses with a fixed baryonic ``Chandrasekhar'' mass of $1.37 \, M_{\odot}$ \citep{nos}.

While there is in principle enough energy to blow away $0.3-0.4 M_{\odot}$ and leave a strange star with $M \leq 1 \, M_{\odot}$, the final outcome depends on many things, including the moment in which the SQM is nucleated inside the collapsing core \citep{Nucleation, German}, the density profile at that time and other factors. We note that, even considering the binding of sub-nuclear origin \cite{BD}, the smallest cores would not produce a final gravitational mass below $1 \, M_{\odot}$ \citep{Di}. Sweeping away a considerable fraction of the mass of outer core seems thus inevitable in a scenario in which the ``strange'' compact object is born in the very supernova event.

It is interesting to consider the case in which a detonation does {\it not} happen, that is, the velocity of a fast deflagration mode would still be a fraction of $c$, meaning that the conversion of the star would take place within $\sim$ \textit{few \, ms}. We can state that the neutrinosphere temperature would be suddenly increased (the SQM hypothesis led to ``reading'' a neutrinosphere temperature of $T_{\nu} \simeq 1.78^{+0.7}_{-0.5}\text{ MeV}$ associated to the SQM energy release in the case of SN 1987A, \cite{Macacos}). At this stage the outer core and envelope are loosely bound, and a suitable gain region for the neutrino flux can produce the explosion with the ejection of $\geq \, 0.3 \, M_{\odot}$ of the former. The suggestion is that a neutrino-driven explosion around $\sim\text{s}$ after prompt shock stalling, due to a refreshed neutrino flux, could be viable. Much in the same way that establishing the basic current supernova picture took several years, a scenario producing a strange star has to be explored with care. 

An important point along the lines is related to the presence of a normal crust. The simple qualitative scenario outlined here would predict a fallback crust of a heavy composition, likely carbon or heavier. The thermal surface emission has to be carefully explored, it may hold spectroscopic clues to confirm or discard this expectation and would be important for a full evaluation.

\section{Conclusions}

The report of a low mass $M = 0.77^{+0.20}_{-0.17} \, M_{\odot}$ for the Central Compact object XMMU J173203.3-344518 born in the supernova HESS J1731-347 is one of the many interesting works seeking for the composition of superdense compact objects in Nature. Its mass is truly unexpected within standard current models of compact stellar remnants, independently of its ultimate composition. On the other hand, an exotic nature of the former is entirely possible, and as we have seen, tentatively consistent with all three direct observations (mass, radius and surface temperature). A summary of these points elaborated above using existing calculations are:  

   \begin{enumerate}
      \item The same quark EoS employed for ``large'' massive objects with reliable determinations of mass and radius naturally accommodates XMMU J173203.3-344518. 
      Other EoS are also capable of explaining the whole set of masses and radius, for example the ones presented by \citet{Ignazio} and \citet{Ferrer2}. The point here is that there is a hint of unification of description, although to pinpoint the ``right'' EoS is still a distant goal.
      
      \item The observations indicating that the surface temperature is quite high, even for the younger age estimate $2-6\text{ kyr}$ \citep{Acero, Cui, Maxted}, exclude rapid cooling within a quark EoS. Moreover, the simplest explanation would be achieved within a ``low gap'' ($\Delta \leq \, 1\text{ MeV}$) complete pairing of quarks. A large value of the gap would be disfavored on these cooling grounds. A new round of detailed calculations will surely shed light on this issue.
         
      \item The idea that strange matter is an ultimate cause of collapse supernovae \citep{Macacos} can be tested in this case. The SQM theory has not produced a supersonic detonation yet, but a very rapid subsonic mode has been obtained. This energy release would need to unbind a large fraction of the outer core to explain the low mass of XMMU J173203.3-344518, and a neutrino-driven wind could be enough for that task.

      \item  We believe that a classification of the supernova event could be attempted using the light echo techniques and spectroscopic observations. A shell of $\sim 2 \, M_{\odot}$ is present around XMMU J173203.3-344518, and remains to be fully characterized. The issue of the whole ejected mass is quite similar to the case of the Crab, for which no definitive identification of a supernova type has been confirmed \citep{AAS}. Given that a so-called ``canonical mass'' has been employed/imposed as certain over the years, and that we now know that there are confirmed lower mass objects \citep{Martinez, nos}, a reanalysis of the Crab case would be in order too.
   \end{enumerate}

   We became aware of a preprint \citep{Di} which shares many of the points made above, not only the structure considerations, but also issues related to the relatively hot temperature reported for the estimated age $2-6\text{ kyr}$ of HESS J1731-347. We hope to have contributed with additional considerations to the case.

Besides confirming all the reported data and analysis, and since the extraction of parameters is always difficult and merits further scrutiny, we can state that the compact object XMMU J173203.3-344518 born in the supernova explosion which left behind the remnant HESS J1731-347 passes the initial consistency checks and opens a window for theoretical and observational work to study its nature and formation history in depth. It is premature to claim any stronger conclusion, although this is an important case and other detections could add to the whole picture. A recent work by \citet{Lin} has reported another candidate with $M = 0.98 \pm \, 0.03 \, M_{\odot}$ that deserves attention in connection to the XMMU J173203.3-344518 case.

\begin{acknowledgements}
This work was supported by the FAPESP Agency (S\~ao Paulo State) under the grant 2020/08518-2 and the CNPq (Federal Government) for the award of a Research Fellowship to JEH. L.M.S. acknowledges CNPq for financial support. The CAPES Agency (Federal Government) and PRG-USP São Paulo are acknowledged for financial support in the form of scholarships. We would like to acknowledge an anonymous referee for valuable scientific suggestions that improved the initial version of the manuscript.

\end{acknowledgements}

% WARNING
%-------------------------------------------------------------------
% Please note that we have included the references to the file aa.dem in
% order to compile it, but we ask you to:
%
% - use BibTeX with the regular commands:
%   \bibliographystyle{aa} % style aa.bst
%   \bibliography{Yourfile} % your references Yourfile.bib
%
% - join the .bib files when you upload your source files
%-------------------------------------------------------------------

\end{document}